\RequirePackage[2020-02-02]{latexrelease}

\documentclass[article,twocolumn,amsmath,amssymb,10pt]{revtex4}
\usepackage[pdftex]{graphicx} 
\usepackage{color}

\begin{document}



\title{Particle charging during pulsed EUV exposures with afterglow effect}

\author{M. Chaudhuri$^{1*}$, L. C. J. Heijmans$^1$, M. van de Kerkhof$^{1,2}$,  P. Krainov$^3$, D. Astakhov$^3$ and A. M. Yakunin$^1$}
\affiliation{$^1$ASML The Netherlands B.V., P.O.Box 324, 5500AH Veldhoven, The Netherlands}
\affiliation{$^2$Department of Applied Physics, Eindhoven University of Technology, PO Box 513, 5600 MB Eindhoven, The Netherlands}
\affiliation{$^3$ISTEQ B.V., Eindhoven, The Netherlands}

\begin{abstract}

The nanoparticle charging processes along with background spatial-temporal plasma profile have been investigated with 3DPIC simulation in a pulsed EUV exposure environment. It is found that the particle charge polarity (positive or negative) strongly depends on its size, location and background transient plasma conditions. The particle (100 nm diameter) charge reaches steady state in a single pulse (20 $\mu$s) within the EUV beam in contrast to particles outside the beam that requires multiple pulses. The larger the particle size, the less number of pulses are required to reach steady state. It is found that the charge of a particle decreases with pressure in a faster rate outside the beam compared to inside. The results are of importance for particle contamination (defectivity) control strategy for EUV lithography machines.     
\end{abstract}

\maketitle

The EUV lithography technology helps the semiconductor industry to follow Moore's law with shrinking transistor sizes and hence higher resolution for high volume manufacturing (HVM) products of 7 nm and 5 nm nodes. This technology uses highly energetic EUV photons (energy $\sim$ 92 eV) with the wavelength of 13.5 nm. One of the side effect of this development is the generation of EUV induced plasma due to the interaction of such highly energetic EUV photons with low pressure (1-10 Pa) background hydrogen gas~\cite{BanineJAP2006, van_der_Horst_2014, Roderik_2018, Kerkhof_2019, Kerkhof_2020, Kerkhof_2021a,Kerkhof_2021}. It is to be noted that the EUV photon induced plasmas differ from traditional low-temperature laboratory plasmas (DC discharge, capacitively coupled radio frequency plasmas, inductively coupled plasma, dielectric barrier discharges, etc), in a sense that there exists no continuous external power supply to sustain the discharge. Instead it is generated by a pulsed EUV source. This creates an EUV pulse of few tens of nanoseconds every 20 $\mu$s. Hence the EUV induced plasma exists mainly in a decaying afterglow phase. Although the photon induced plasmas is an active subject of investigation for astrophysical plasmas for some time~\cite{Goertz1989,Rosenberg1995,Fortov1998,PhysRevLett.84.6034,2001_Weigartner}, the experimental research has gained momentum recently due to industrial applications in the form of EUV lithography. 

During such lithography manufacturing processes, sub-micron particles must be controlled and kept away from critical surfaces to be imaged, exposed or measured such as photomasks and wafers. The force balance on these particles, is often dominated by coulomb forces where particle charge plays an important role. The transient charging process strongly depends on the location and size of the particles. Typically, the particle charge in plasma environment is determined by the flux balance of different plasma components, primarily electrons and ions. This is also true for EUV induced plasma region. However, within the EUV beam, the photons also play important role for charging process. The strong coupling between the local transient plasma parameters and the particle charge also carries on in the post-discharge phase (afterglow). The residual electric charges of microparticles in the afterglow were observed for the first time under microgravity conditions~\cite{Ivlev2003}. Systematic investigations by Cou{\"e}del et al. inside a capacitively-coupled RF discharge with nanoparticles show that decharging processes and residual charges are strongly linked to the diffusion of electrons and ions in the post discharge  
\cite{Couedel2006,Couedel2008,Couedel2008b,Couedel2009a,Layden2011a}. The afterglow physics have been explored with theoretical investigations by Denysenko et al. which reveal that the presence of metastable atoms can considerably influence the residual electric charges of the microparticles \cite{Denysenko2011,Denysenko2013}. 
Several other experiments have also been performed to understand the afterglow physics on particle charging, such as, effect of an external electric field on microparticle charge distribution~\cite{Woerner2013}, spontaneous dust pulse formation under microgravity~\cite{chaudhuri_2021}, effect of initial particle cloud configuration/shape on cloud dynamics (simple Coulomb expansion to more complex behavior such as Coulomb fission)~\cite{Meyer2016,Merlino2016}. In recent times dust response to the afterglow (both temporal and spatial) plasma has gained attention~\cite{Minderhout2019, Minderhout2020, Minderhout2021, Denysenko2021, Chaubey2021, Suresh2021, chaudhuri_2022, Chaubey2022, Staps2022, Couedel2022, Huijstee2022}. It is to be noted that in case of 20 $\mu$s pulsed EUV environment in lithography machines, the afterglow physics with non-LTE (Local Thermodynamic Equilibrium) conditions is fundamentally different than those mentioned above. The relevant defectivity issues are associated with plasma-particle interaction at the single particle level rather than particle cloud effects. {\it The goal of this work is to present the current results of single particle volume charging processes in transient EUV exposure environment focussing on afterglow effects.}   

\begin{figure}
\includegraphics[width=0.95\linewidth]{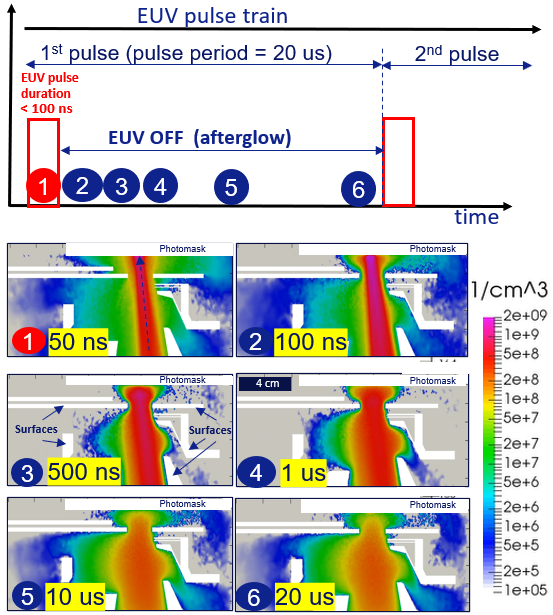}
\caption{The single EUV pulse is 20 $\mu$s out of which EUV power is ON for $<$ 100 ns. Spatial distribution profile of electron density at Reticle Mini-Environment (RME) is shown within the EUV pulse ($<$ 100 ns) and after pulse (100 ns - 20 $\mu$s) in the afterglow condition. Within the EUV-pulse, the arrow indicates the EUV-beam path. It is clear that the spatial electron desnity profiles change with time in the afterglow condition. The brief description of RME within ASML lithography machines can be found in Reference~\cite{Kerkhof_2021a} } 
\label{transient_plasma_variation}  
\end{figure}

\begin{figure}
\includegraphics[width=0.95\linewidth]{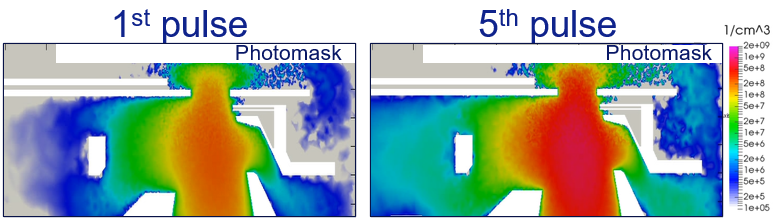}
\caption{EUV induced afterglow in the RME at the end of 1st and 5th pulse . The plasma accumulation over 5 pulses is clearly visible.}
\label{plasma_build_up}  
\end{figure}

The afterglow physics plays important role to understand particle contamination control in EUV lithographic machines under pulsed mode operation. The spatial and temporal evolution of the EUV plasmas has been investigated experimentally and with particle-in-cell (PIC) model~\cite{van_der_Horst_2016, Astakhov_2016,Beckers_2016,van_de_ven_2018}. The plasma evolution occurs in different stages: at first the plasma creation happens due to direct photoionization of the hydrogen gas (ionization energy of 15.4 eV) during the passage of highly energetic EUV photons. In this phase, the plasma contains highly energetic electrons with excess photon energy and non-Maxwellian energy distribution. A part of these electrons then move towards the wall leaving behind a positive space charge region which confines the remaining electrons and allow positive ions to accelerate towards walls. The confined electrons lose their energy within few tens of nanoseconds due to electron impact ionization and increase the plasma density. After this stage, plasma starts to expand and the local electron density decreases rapidly. During the last stage of the process, the electrons are supposed to reach their equilibrium temperature with plasma density continues to decrease due to ambipolar diffusion and recombination processes at wall. The energy is supplied within few tens of nanoseconds by the EUV pulse while it takes $\sim$ 20 $\mu$s for the plasma to completely extinguish which makes the highly transient plasmas. Such spatial-temporal variations of electron density within a single pulse at the RME region is shown in Figure~\ref{transient_plasma_variation}. The afterglow spatial electron density variation at the end of 1st and 5th pulse is shown in Figure~\ref{plasma_build_up} which clearly shows the plasma build up over the pulses.

The charge of a particle is an important parameter for particle dynamics including volume and surface forces. Depending on the locations (inside or outside of EUV beam) and time (during or after the EUV pulse), the particles are charged through a balance between electron, ion and photon fluxes. The charging equation can be written as,
\begin{equation}
\label{electron_flux} 
\frac{dZ}{dt} = \sum I_\alpha
\end{equation}
Here $\alpha$ reprsents different plasma components, electrons (e), ions (i) or photons (ph).

The orbital motion limited (OML) theory by Mott-Smith and Langmuir ~\cite{langmuir_1926, allen_1992,kennedy_allen_2003} has long been a standard tool to evaluate electron and ion fluxes. However, in plasma conditions, the parameters of an electron energy distribution changes significantly starting from a strongly non-Maxwellian distribution (electron energy $\sim$ 80 eV) to nearly thermal plasma (electron energy $\sim$ 1 eV). In this scenario, the pulse average method is not applicable and a detailed analysis with electron energy distribution function $f(E_e)$ is needed. The electron flux expression can be written as:

\begin{equation}
\label{electron_flux} 
\begin{aligned}
I_e = \pi a^2 n_e  \times 
    \int\sqrt\frac{2E_e}{m_e}(1-Y(\phi_p, E_e))f_e(E_e) \\ 
\times \left(1 + \frac{e\phi_p}{E_e}\right)\Theta(E_e + e\phi_p) d E_e.
\end{aligned}
\end{equation}
Here, $I_{e},  n_{e}, E_e, m_e$ are the flux, density, energy and mass for electrons respectively. The parameter $f_e(E)$ is the electron energy distribution function, $Y$ is the total yield of secondary electrons and $\Theta$ is a step function. The floating potential of the particle is represented as $\phi_p$ and $a$ is radius of the particle. In contrast, the ion flux can be written asssuming a thermal distribution of ions with a temperature $T_i$, 
\begin{equation}
\label{ion_flux} 
I_i = \pi a^2\sum_i n_i(q_i/e) v_{Ti} \times \begin{cases}
    \exp(-e\phi_p/kT_i), & \text{if $\phi_p \ge 0$}.\\
    (1 - e\phi_p/kT_i), & \text{if $\phi_p < 0$}.
  \end{cases}
\end{equation}
Here, $I_i,  n_i$ are the flux and density for particular ions respectively. $k$ is the Boltzmann constant and $e$ is the elementary charge. In EUV induced hydrogen plasma, the ion populations are different at different time scales. Initially within the EUV pulse ($\sim$100 ns), the $H_2^+$ ion population is maximum. But after the pulse most of the $H_2^+$ ions start to transform to $H_3^+$ ions. As a result, the $H_2^+$ ion density decreases rapidly after the pulse and the $H_3^+$ ion density increases. Two distinct decay zones can be identified for $H_2^+$: fast decay (upto $\sim$ 250 ns) and slow decay for the rest of the period. The $H_3^+$ ion density reaches the maximum value at $\sim$ 700 ns when most of the $H_2^+$$\rightarrow$ $H_3^+$ conversion is done. In this decay period the $H^+$ ion density decreases very slowly. In all these temporal regimes, the quasineutrality condition ($n_e = \sum_i n_i$) in the bulk locations is applicable.

Similarly, the photon flux to the particle can be written as, 
\begin{equation}
\label{photoelectron_flux} 
I_{ph} = \pi a^2 \gamma(\phi_p) \frac{I}{h\nu}
\end{equation}
Here, $\gamma$ is the photoionization yield (mean number of electrons expelled due to single photon absorption), $I$ is the local EUV intensity at the particle. The correction of a yield can be done by following factor,  
\begin{equation}
\label{photoelectron_flux_correction} 
\gamma(\phi_p) = \gamma_0\int_{e\phi_p}^{+\infty}f_{ph}(E)dE
\end{equation}
Here, $\gamma_0$ is the yield in absence of charge on the particle and $f_{ph}$ is the energy distribution function of the emitted electrons normalized by one.

\begin{figure}
\includegraphics[width=0.95\linewidth]{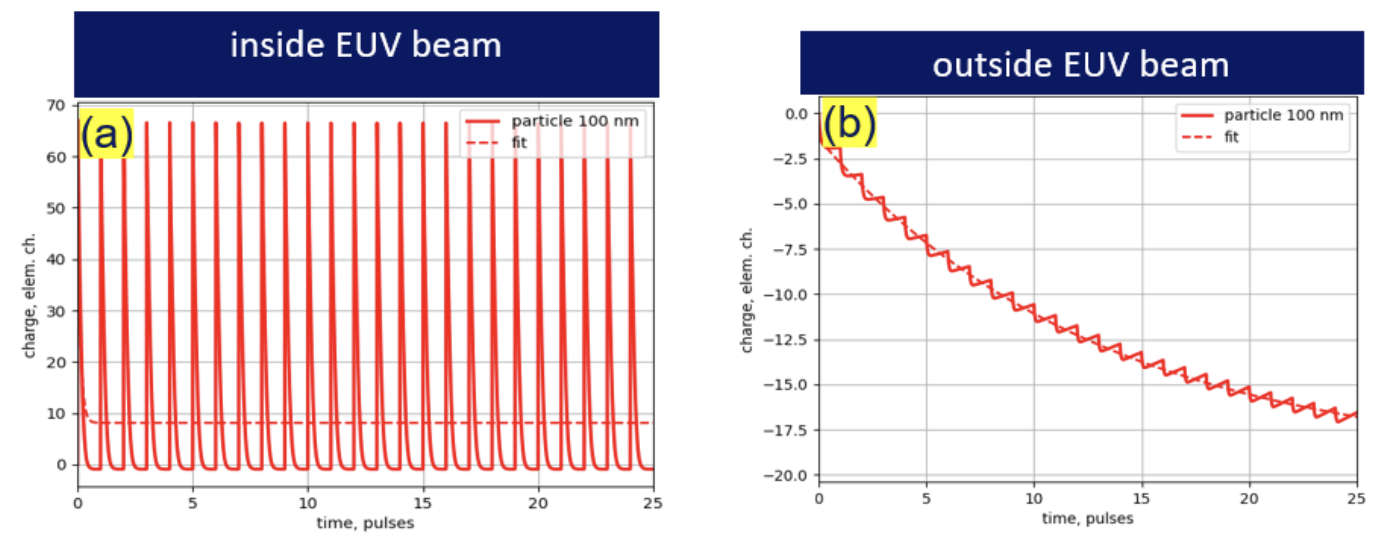}
\caption{(a) The time-dependednt charge of a 100 nm particle inside the EUV beam. The particle charge reaches steady state in a single pulse and is always positive due to photo-electric effect (see text) (b) The charge variation of the same particle outside EUV beam. In this case, the charge is always negative and multiple pulses are required to reach steady state.}
\label{transient_charging_inside_outside}  
\end{figure}

Within the first 100 ns of each pulse duration of 20 $\mu$s, the EUV photons charge particles positive within the beam by expelling electrons from it (photo-electric effect). The afterglow effect starts when the EUV power is OFF and the particle charge reduces as in this regime the electron flux dominates. The combination of these effects cause a periodic oscillation of the charge on a particle. The multi-pulse transient charging inside the EUV beam region is shown in Figure~\ref{transient_charging_inside_outside}a. The particle charges positive during the EUV-pulse due to the photons and negative between pulses due to the EUV-induced plasma. Particles outside of the EUV beam in the bulk have the traditional charging mechanisms with a balance between electron and ion fluxes making it negative due to higher mobility of electrons. In that case, the particle charge at a particular location in the bulk only slightly oscillates with the EUV pulses, due to the changing (transient) plasma parameters as illustrated for a 100 nm particle in Figure~\ref{transient_charging_inside_outside}b and the same particle acquires negative steady state charge through multiple pulses. In both cases, the dynamics of the particle charge was fitted as $Q(t) = Q_s + Q_0 \times \exp(-t/t_{ch})$ . Here $Q_s$ is the stationary charge, $(Q_s + Q_0 = 0)$ is the initial charge and $t_{ch}$ is the characteristic charging time. The fitted curves are shown by the dashed line. In the vicinity of the surfaces, the photoelectric effect of the surface may dominate and significantly contribute to the incoming electron flux to the particle. 

\begin{figure}
\includegraphics[width=0.95\linewidth]{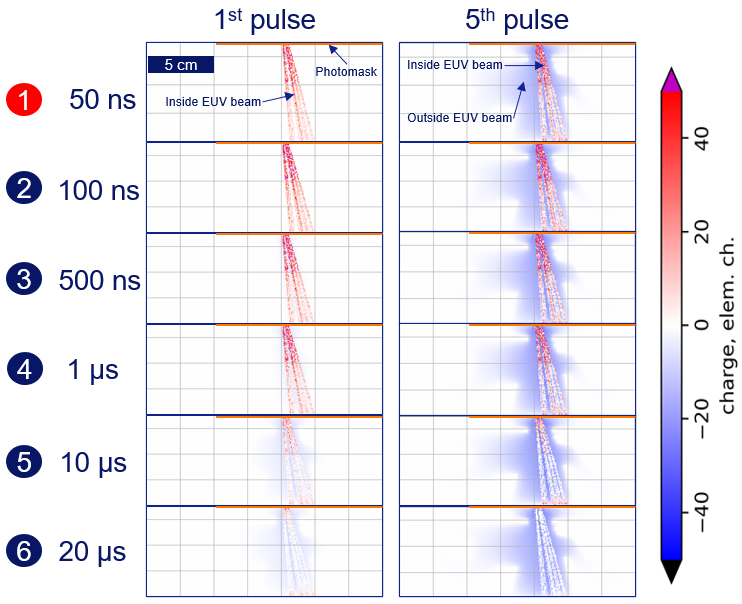}
\caption{The particle charge distribution at RME for 1st (left) and 5th (right) EUV pulses at different time scales. The negative charge is represented by blue and positive charge by red colours. At the end of each pulse, the particles within the beam location becomes almost neutral. The time sequence 1-6 corresponds to that shown in Figure~\ref{transient_plasma_variation}}
\label{charge_distribution}  
\end{figure}

The particle charge for 100 nm particles at different locations and times are shown in Figure~\ref{charge_distribution}. The figure shows that the  particle charge varies throughout the volume. Particles are negative in most of the volume, but get positively charged inside the beam area due to the photo-ionisation. The charging time is much shorter inside the beam than outside the beam; as anticipated in Figure~\ref{transient_charging_inside_outside}. This implies that the particle charge is influenced by its trajectory-history, especially for particles outside the beam, as they might not be in equilibrium with their surrounding yet. The charging mechanisms depends strongly on particle size. Considering OML theory, it can be shown that $dQ/dt \sim a^2$ where $a$ is the radius of particle. On the other hand, the equilibrium potential depends only on plasma conditions, not particle size so that $Q_s \sim a$. Therefore, the time required to come to this charge scales as $t_{ch} \sim Q_s/(dQ/dt) \sim 1/a$. This implies that faster charging occurs for bigger particles. 

\begin{figure}
\includegraphics[width=0.85\linewidth]{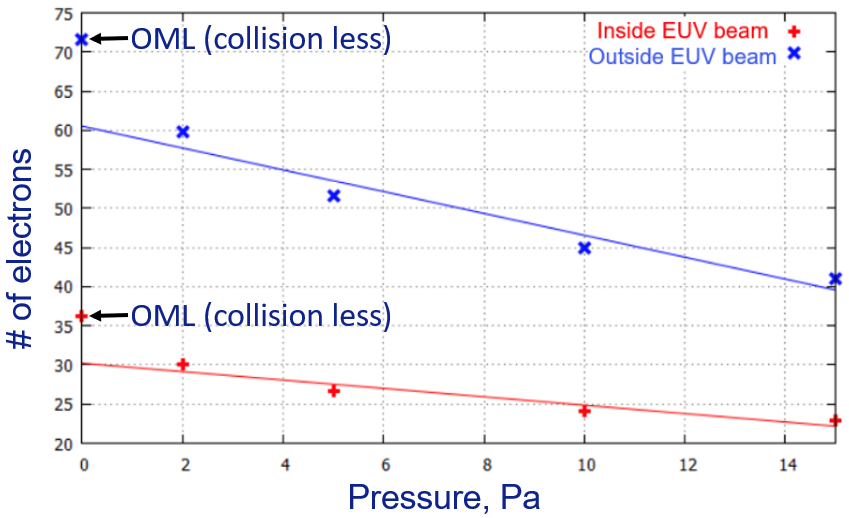}
\caption{Steady state charge variation of 100 nm particle with pressure for both inside and outside of EUV beam. The ion-neutral collisionality significantly reduced particle charge compared to collisionless (OML) cases. The slope of the curve is represented by s-coefficient. The variations of s-coefficient at different locations (inside and outside the EUV beam) are shown in Figure~\ref{charging_slope_different_zones}.} 
\label{charging_slope}  
\end{figure}

\begin{figure}[h]
\includegraphics[width=0.85\linewidth]{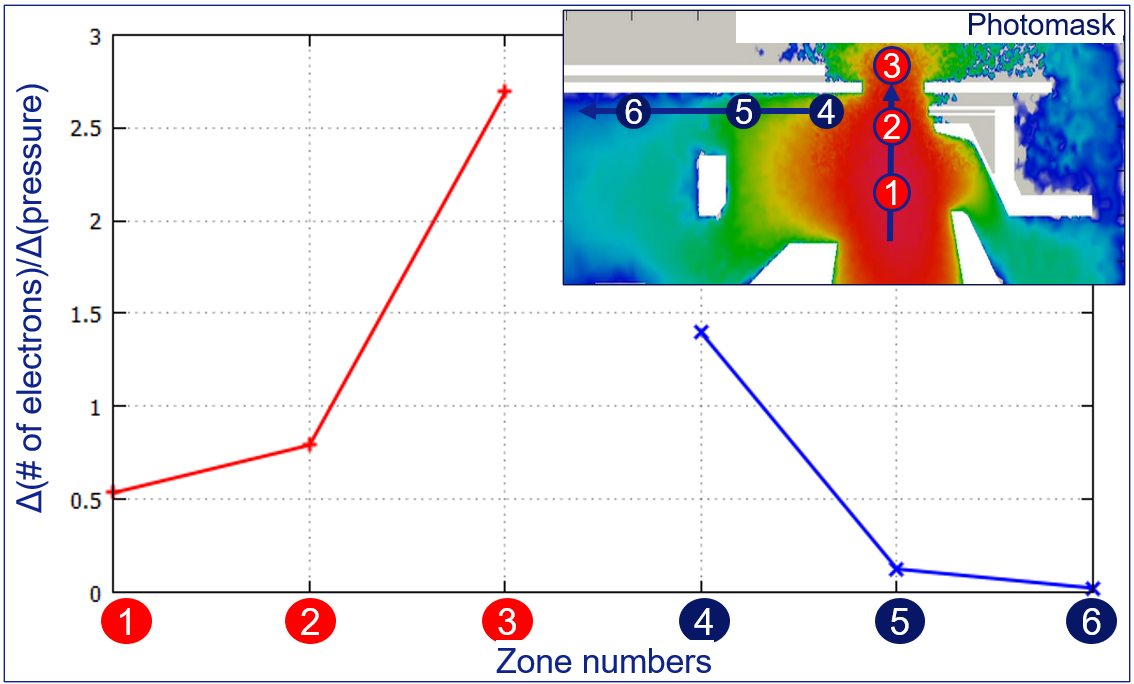}
\caption{The pressure sensitivity of 100 nm particle charge has been estimated at different locations inside (red curve) and outside (blue curve) of EUV beam. The zone numbers are shown in the inset.} 
\label{charging_slope_different_zones}  
\end{figure}

It is to be noted that the above mentioned orbital motion limited (OML) theory is commonly used to estimate particle charge which assumes collisionless, ballistic ion and electron trajectories in the vicinity of an isolated particle without the presence of any kind of interaction potential barrier~\cite{allen_1992, kennedy_allen_2003}. In the last decade, however, it was recognized that ion-neutral collisions (neglected in the OML theory) can in fact represent a very important factor affecting the ion flux collected by the particle. The ion angular momentum is lost due to collisions with neutrals and this effect significantly enhances ion flux to the particle and therefore the charge of the particle decreases. In this regime, the collision enhanced ion model should be used, $I_i \approx \sqrt{8\pi}a^2n_iv_{\rm Ti}(1 + z\tau + 0.1z^2\tau^2\eta)$ where $z = Ze^2/aT_e$ is the normalized particle charge, $\tau = T_e/T_i$ is the electron-to-ion temperature ratio and $\eta = \lambda_{\rm D}/\ell_i$ is the ion collisionality index (ratio of effective Debye length and ion mean free path). This has been evidenced in molecular dynamics ~\cite{zobnin_2000} and particle-in-cell ~\cite{hutchinson_2007} simulations, semi-analytic kinetic theory calculations ~\cite{lampe_2003, Zobnin_2008} and a number of experiments (e.g., in refs. ~\cite{PhysRevLett.93.085001.Ratynskaia, PhysRevE.72.016406.Khrapak, Khrapak_2012, Antonova_2019}). For a complete reference list see refs.~\cite{Fortov2005PR, Khrapak_2009}). 
This effect has been investigated (inside and outside the EUV beam) considering steady state charge over multi-pulse scenario. It is found that pressure sensitivity of particle charge outside of EUV beam is higher than that inside the EUV beam as shown in Figure~\ref{charging_slope}. The computational result shows that the charge values in weakly collisional regime are significantly less than that calculated in collisionless regime using OML theory and consistent with previous experimental results~\cite{PhysRevLett.93.085001.Ratynskaia, PhysRevE.72.016406.Khrapak,Khrapak_2012,Antonova_2019}. The slope of the pressure sensitivity curve shows the dependence of particle charge on pressure, the higher the slope the faster is the charge reduction with increasing pressure. It is found that outside EUV beam, the charge reduction with pressure is much faster close to the beam compared the far distance region. Inside the beam, charge reduction with pressure is faster at the region closer to the reticle. These dependencies at different regions are shown in Figure~\ref{charging_slope_different_zones}. An experimental results of transient particle charge measurements in EUV time scale is needed for direct comparison with simulation results and is kept for our future work.\\

In conclusion, it is found that particles charge strongly depends on their locations inside and outside EUV beam. Inside EUV beam the particles become positively charged due to photoionization  whereas outside the EUV beam the particles are negatively charged. The particles acquire steady state charge within EUV but it takes many pulses outside the EUV beam. The particles of different sizes acquire steady state charge depending on number of pulses. The charge of the particle decreases with increasing neutral gas pressure and it depends on the location of the particle. It is also important to perform dedicated experiments for charge measurements in afterglow EUV plasma and compare with 3DPIC simulation data as mentioned in this work. This has been kept as future work.


* Corresponding author: manis.chaudhuri@asml.com

\end{document}